\documentclass{aa}
\usepackage{graphicx}

\begin{document}  

\thesaurus{ ; ; ; ; ; }

\title{Kinematics of Molecular Gas in the Nucleus of NGC\,1068,  
from H$_{2}$ line emission observed with VLT
\thanks{ Based on observations collected at the ESO/Paranal ANTU telescope, 
Proposal 63.P-0167A.}} 
\author{ D.\,Alloin \inst{1}, E.\,Galliano \inst{1}, J.G.\,Cuby 
\inst{1}, O.\,Marco \inst{1}, D.\,Rouan \inst{2}, Y.\,Cl\'enet 
\inst{2}, G.L.\,Granato \inst{3} and A.\,Franceschini \inst{3}}
\institute{ European Southern Observatory, Casilla 19001, 
Santiago, Chile
\and  Observatoire Paris, Meudon, F-92190, France
\and  Osservatorio Astronomico di Padova, I-35122 Padova, Italy}
\offprints{D. Alloin} 
\mail{dalloin@eso.org}
\date{Submitted to Astronomy \& Astrophysics Letters}
\titlerunning{Kinematics of Molecular Gas in NGC\,1068, from H$_{2}$ line 
emission}
\authorrunning{D.Alloin \& al}
\maketitle

\begin{abstract}

We present results about the distribution and kinematics of the 
molecular environment of 
the AGN in NGC\,1068, over a $1.5'' \times 3.5''$ region around the central 
engine in NGC\,1068, derived from H$_{2}$ line emission detected with 
ISAAC at VLT/ANTU on ESO/Paranal. The H$_{2}$ emitting molecular gas
is found to be distributed along the East-West direction and with two 
main peak emission (knots) located at a distance of about 70 pc from the 
central engine. The eastern H$_{2}$ knot is more intense than 
the western one.
The line profiles mapped across the entire $1.5'' \times 3.5''$ region, at a 
spatial resolution of $0.3''\times 0.45''$,
appear to be quite complex with either a blue or red wing. At first order, 
we find a velocity difference of 
140 km/s between
the two knots; if interpreted as quasi-keplerian velocity, this
implies a central enclosed mass of $\rm 10^8~M_{\odot}$.  


\end{abstract}

\section{Introduction}

Given the proximity of NGC\,1068 (14.4 Mpc and corresponding scale 
of 70 pc per $1''$) and the predicted size of the molecular 
torus in an Active Galactic Nucleus (AGN) --from 1 to 100 pc--, any 
trace or signature of a molecular torus in NGC\,1068 must be searched from 
data collected under sub-arcsec image quality. Several recent 
discoveries point 
towards the presence of a conspicuous and structured molecular/dusty 
environment around the central engine of NGC1068 (e.g. Gallimore et al 
1997, Rouan et al 1998, Marco \& Alloin 2000, Schinnerer et al 2000, 
and references given in these papers). One piece of information still 
missing to ascertain the existence of a rotating torus --in addition 
to other possible molecular components--, is the kinematical status of 
the molecular/dusty material. Such information is available for the cold 
molecular gas, from 
recent interferometric work in the CO(2--1) line (Schinnerer et al 2000):
the authors infer the presence of a 
warped disc of cold molecular gas.
In this study we have chosen to probe the molecular gas through 
the H$_{2}$ 1-0 S(1) line (rest $\rm \lambda =  2.12 \mu m$). Being related 
to the hot
molecular gas and having specific excitation mechanisms, it provides 
complementary information to that derived from the CO transition which
traces the cold molecular gas. After the pioneering discovery of
the H$_{2}$ 2.12 $\mu m$ line emission in NGC\,1068 by Thompson et al 
(1978), a first attempt to image 
the AGN of NGC\,1068 in the H$_{2}$ 2.12 $\mu m$ line has been 
reported by Rotaciuc et al (1991), covering a $10'' \times 10''$ region 
at a resolution of $\approx 1''$. It is imperative to push further the 
spatial resolution.

Adaptive optics high resolution K, L and M band images of the AGN
have unveiled the presence and structure of hot to warm dust (Rouan et al 
1998, Marco \& Alloin 2000)
 within the $1'' \times 1''$ region around 
the central engine. We have selected from these observations two particular 
directions for our
kinematical study: (a) PA $=$ 102\degr\ possibly tracing the
equatorial plane of the molecular/dusty torus around the central engine;
(b) PA $=$ 12\degr\ which features the axis of the torus and is found to
be close to the axis of the ionizing cone to the North-East.
The location of the central engine -- only visible directly in the 
IR as an unresolved core carrying around 90\% of the emission and in 
the radio as the radio source S1 in Muxlow et al (1996) -- is taken here
as  that derived 
by Marco et al (1997) from simultaneous K and I band high resolution 
imaging with adaptive optics. Given the error bar on this position
 ($\rm \pm 0.15''$) , it is coincident with the astrometric position of the 
 12.4 $\mu$m unresolved
 core (Braatz et al 1993), the K band peak observed by Thatte et al (1997) and 
 the 
center of symmetry of the polarization pattern in the near-IR and mid-IR 
(Lumsden 
et al 1999). We leave aside attempts at locating the central engine 
from UV data which are quite sensitive to dust extinction and
provide so far discrepant results (Capetti et al 1995a \& b, Kishimoto 1999). 

We present and discuss results of long-slit spectroscopy in the 
near-IR, obtained 
with ISAAC at VLT/ANTU on ESO/Paranal. In this Letter, we concentrate
on the H$_{2}$ 2.12 $\mu m$ line emission observed over a 
$1.5'' \times 3.5''$ region through the central 
engine at PA $=$ 102\degr. The kinematics of the hot molecular 
gas in particular is investigated
to test the presence of an eventual super massive object 
in NGC\,1068.

\section{ Data collection and Reduction} 

The observations were performed using the SWS1 short wavelength arm of 
the instrument ISAAC attached to the Nasmyth focus of ANTU (Moorwood et al 
1999,
 Cuby et al 2000). The measured seeing value was of  $0.5''$ (FWHM) in the 
 K band. 
 The slit width was set to $0.3''$ while its length
was of $2'$ at PA $=$ 102\degr. A spectrum was first obtained with 
the slit centered on the near-IR unresolved core, imaged prior to the 
spectroscopic observations. Then the position of the slit was offset by 
$0.3''$ and $0.6''$ on each side of the core, to the North 
and to the South, providing a complete mapping of a $1.5'' \times 1'$ area. 
A complete description of the data collection and reduction procedures is
given in Galliano \& Alloin (2001). The final spectral resolution at the 
observed
 2.15 $\mu m$ wavelength of the H$_{2}$ line is 35 km/s. \\
From the reduced 2D spectra,  we extracted a series of 1D spectra, through a 
window 
3 pixels-high (i.e.
$0.45''$) along the slit, and with a sliding step of 1 pixel
($0.15''$). In the very central area, the intense continuum produces a 
fringing pattern at 
the level of 4\% (peak to peak) 
that cannot be fully corrected, leaving some residuals which prevent the 
measurement of eventual faint H$_{2}$ line emission at a distance less than 
15 pc from the central engine.

\section{Results and Discussion}

\subsection{H$_{2}$ line profiles}

We provide in Figure~\ref{fig1}, a set of the H$_{2}$ line profiles observed at
various positions across the $1.5'' \times 3.5''$ central area extended 
along the equatorial plane 
of the suspected torus (PA $=$ 102\degr).
At each position, the displayed line profile corresponds to 
an $0.3'' \times 0.45''$ 
emitting patch. We did not 
deconvolve the individual successive spectra: given the seeing 
($0.5''$ in 
the K band), each spectrum is only moderately contaminated by its 
neighbours.
The H$_{2}$ line profiles are presented in the following manner: for 
each frame, offset coordinates
of the patch to which the line-profile refers are given
with respect to the central engine location. The horizontal axis provides 
the offsets
along PA$=$ 102\degr~and the vertical axis the offsets along PA$=$ 12\degr.
In each frame, the black (thin) line features the line profile observed to 
the West of the AGN (from $-1.5''$ to $-0.3''$), while the 
red (thick) line features the line profile observed to 
the East (from $+0.3''$ to $+1.5''$). Therefore, each frame provides the 
line profile 
of a given patch and the line profile of its symmetrical counter part 
with respect
 to the axis of the suspected torus.
Because of a large difference between the $\rm H_{2}$ line fluxes to the 
East and West of the central engine, we have rescaled by a factor 3, in Figure~\ref{fig1},
the line profiles detected to the West.
In most of the frames, the $\rm H_{2}$ line shows a complex profile: from 
double peak (North) 
to the presence of an extended wing ( blue wing to the West and red wing to 
the East). 
The velocity difference between the main peaks in two symmetrical patches is 
maximum for patches [$0''$;$0.9''$] and [$0''$;$-0.9''$] and reaches a value 
of $\rm \Delta V = 140~km/s$.

\subsection{Spatial distribution of the H$_{2}$ line emission}

For each $0.3'' \times 0.45''$  emitting patch we have measured the total 
integrated H$_{2}$ line emission and,
from these values, we have reconstructed the 2D map in H$_{2}$ line
emission shown in Figure~\ref{fig2}. The existence of two symmetrical emission knots 
(eastern and western knots) with large flux
difference is conspicuous. This map can be compared to the early image
in the H$_{2}$ line provided by Rotaciuc et al (1991, their figure 2), 
within the limitation 
of different
spatial resolutions and spatial coverage. Indeed, the ISAAC reconstructed 
H$_{2}$ map is consistent 
with the inner part of the image by Rotaciuc, although the more limited area
covered by the ISAAC data set may miss the maximum of the western knot (which 
is also slightly shifted to the South). A detailed flux comparison between 
H$_{2}$
data sets obtained previously (references in Section 1) and the ISAAC data
set is presented in Galliano \& Alloin (2001). We concentrate here on the 
positional/kinematical aspects measured for the first
time from this ISAAC H$_{2}$ data. 
The strong eastern H$_{2}$ knot, located about $1''$ to the East
of the central engine, is also well identified in velocity 
space (Figure~\ref{diagposvit_0}). 
In addition to the eastern and western knots, extended and asymmetric wings 
on the line profiles 
suggest the presence of an extended source of H$_{2}$ emission, possibly 
in the form 
of an outflow (Galliano \& Alloin 2001). 
We have also reconstructed a 2D map of the average continuum between 
$2.1{\mu}m$ and $2.2{\mu}m$ 
(not shown here) from which we have
positioned the cross featuring the central engine in Figure 2.

\subsection{Kinematics of the H$_{2}$ molecular material}

\begin{figure}

\mbox{\resizebox{\hsize}{!}{\includegraphics{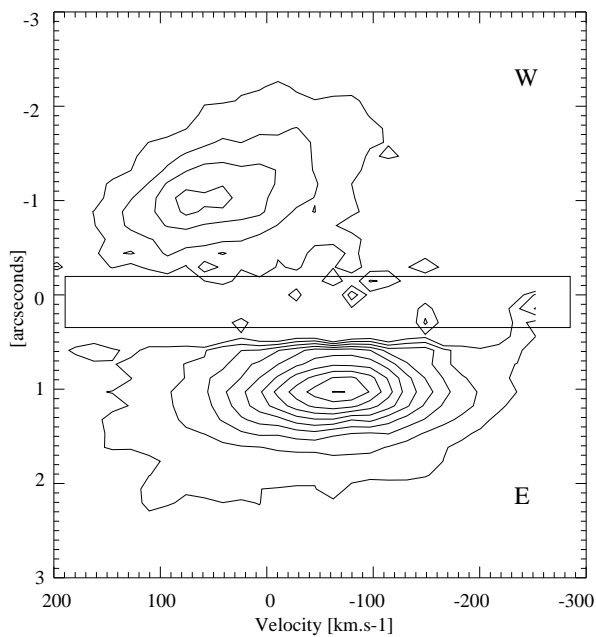}}}

\caption{Position--Velocity diagram for the spectrum at PA$=$102\degr~through 
the central engine. The continuum was fitted and substracted. The contours 
are at 10\%, 20\%,..., 100\% of the eastern peak. Inside the central marked box, 
the 
residual fringing pattern has been 
removed for 
clarity of the diagram.  Outside 
the box, 
the fringing is absent.} 

\label{diagposvit_0}

\end{figure}

\begin{figure*}
\resizebox{12cm}{!}{\includegraphics*[scale=1.]{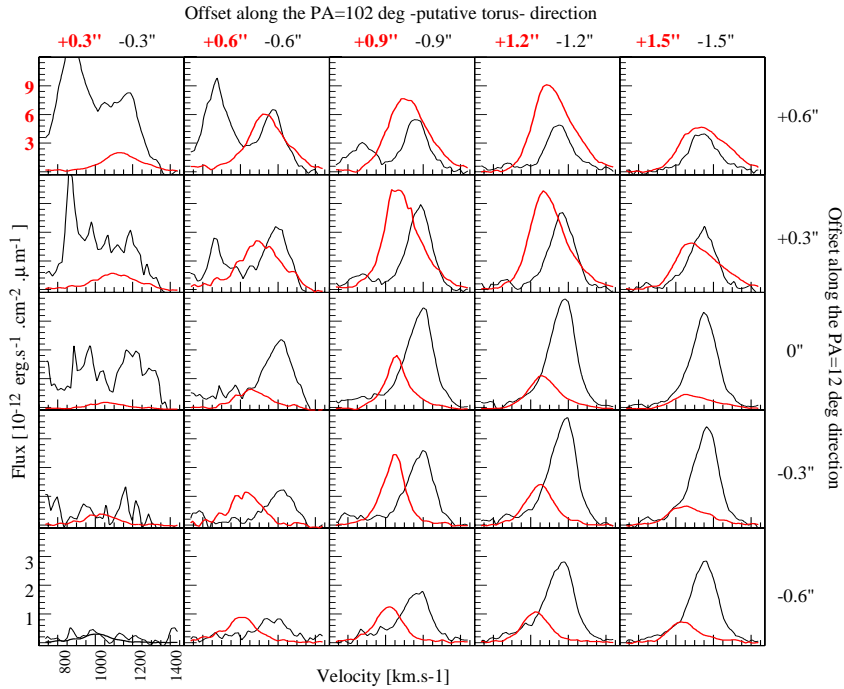}}
\hfill
\parbox[b]{55mm}{
\caption{The H$_{2}$ line-profile data: each frame corresponds to an
$0.3'' \times 0.45''$ emitting area. The center of the area is offset from the 
central engine position [$0'' \times 0''$] by the quantities provided along
the right side of the Y-axis (offset along PA $=$ 102 \degr) and the upper 
X-axis (offset along PA $=$ 12\degr ), the X-axis offset being positive to 
the East and negative to 
the West, while the Y-axis offset is positive to the North and negative 
to the 
South. The lower-left frame provides as well the velocity scale applicable 
to
all frames. The thin (black) line refers to the West of the AGN,
while the
thick (red) line refers to the East. As the first aim is to 
show the velocity shifts, the western profiles have been rescaled in flux 
by a factor 3. The flux scale is given on the left side of the Y-axis: 
the upper (red) one refers to the eastern profiles, the lower 
(black one) to the western ones.} 
\label{fig1}}
\end{figure*}

\begin{figure*}
\resizebox{12cm}{!}{\includegraphics[scale=1.]
{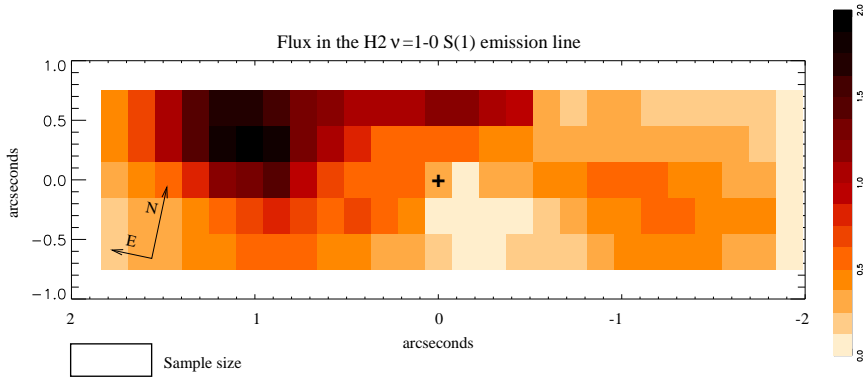}}
\hfill
\parbox[b]{55mm}{
\caption {Restored map of the $1.5'' \times 3.5''$ central area, in the 
H$_{2}$ 1.12 $\mu$m integrated line emission. The intensity scale is
linear and reaches $\rm 2 \times 10^{-14} erg.s^{-1}.cm^{-2}$. The cross 
locates the maximum of the continuum emission as found from this ISAAC data
set and features the location of the central engine as derived from K
 band adaptive optics measurements by Marco 
et al (1997).}
\label{fig2}}
\end{figure*}

As a first order result, a jump in radial velocity of about 
$\rm 140 ~(\pm 5) ~km/s$ 
is detected between the eastern and western knots at 70 pc from 
the central engine. The [Position--
Velocity] diagram (Figure~\ref{diagposvit_0}) enlights this result. If the material
associated with these knots pertains to a structured 
ensemble of molecular
clouds in keplerian orbit, this implies the presence of an enclosed object 
with dynamical mass $\rm 10^8~M_{\odot}$. A comparable value is found 
by Schinnerer et al (2000) from CO line emission observed on a scale of 
$\rm \pm 70 pc$. Yet, the enclosed central mass derived
from H$_{2}$O maser observations on a scale of less than 1 pc, is only 
$\rm 1.5 \times 10^7~M_{\odot}$ (Greenhill \& Gwinn 1997).
It should be noticed as well that the rotation axis of the molecular 
and maser ``pseudo'' discs differ by at least 30\degr. This result suggests 
that the mass enclosed within the 70 pc radius region around the central
engine in NGC\,1068 is shared among different components.
 
At second order, across the western knot itself, an increase of the H$_{2}$ 
line velocity is observed towards the
center, as expected in the case of pseudo-keplerian rotation of the system,
but a similar behaviour does not appear in an obvious way in the eastern knot 
where 
the brightest peak 
seems to remain at constant blueshift when approaching the center. Considering 
the 
large flux difference between both sides, we cannot exclude a superposition 
of 
components along the line of sight to the East of the central engine, 
or the existence of 
peculiar kinematics as discussed from CO line data by Schinnerer et 
al (2000).
A more refined analysis of the velocity field of the H$_{2}$ emitting 
molecular 
material is in progress,
aiming in particular at fitting the line profiles across the entire region.

\subsection{Comparison with CO high resolution map and Discussion}
 
The integrated H$_{2}$ line map derived from ISAAC data can be compared 
with the $\rm ^{12}CO(2-1)$ map recently obtained at IRAM with an 
$0.7'' \times 0.7''$ beam 
by Schinnerer et al (2000). The strong
H$_{2}$ line emission of the eastern knot is detected as well in
CO, at the same location. On the western side, the ISAAC spatial
coverage does not allow to check whether the CO peak -- located $1.5''$ 
to the West 
and $1''$ to the South --  is also emitting in the H$_{2}$ line. However, 
we do 
detect in H$_{2}$
the tip of the western CO knot.
The spatial coincidence of CO and H$_{2}$ emission in the eastern and 
western knots is interesting and 
indicative that, in these knots, H$_{2}$ is emitted in photo (or/and X-ray)
dissociation 
regions (PDR or XDR) 
at the surface of molecular clouds. Given the AGN environment, rich
in UV photons, X-rays and shocks, it is still unclear what would be the
main source of excitation of the H$_{2}$ line. For the time-being, let us 
simply assume that the H$_{2}$ line emission originates from molecular gas 
at temperature around 2000 K (e.g. Hawarden et al 1995):
 we find that the total observed H$_{2}$ line emission in 
the $\rm 3.5''\times 1.5''$ central area is of 
$\rm (20 \pm 6).10^{-14}~ erg.s^{-1}.cm^{-2}$, corresponding to a total 
luminosity of $\rm 1.3 \times 10^{6}~ L_{\odot}$. 
Following 
Thompson (1978), an amount of $\rm 3500~ M_{\odot}$ of hot 
molecular gas is needed to produce this emission. Conversely, CO line 
emission is 
thought to arise from cold molecular gas with temperature at least 
one order of magnitude less (e.g. Barvainis et al, 1997). 
The total mass of molecular hydrogen derived from the CO line flux by 
Schinnerer \& al (2000), over the same region, is of the order of 
$\rm 10^{7}~ M_{\odot}$, 4 orders of magnitude larger than the hot emitting 
H$_{2}$ molecular gas. Such a result is not surprising and is consistent with
the PDRs or XDRs origin of H$_{2}$. Still to be understood is the more
diffuse H$_{2}$ emission which is traced in the wings in the H$_{2}$ line
profiles (Figure 1) and we are developping a kinematical model for this 
extended component (Galliano \& Alloin 2001). 

Another interesting point to resolve is whether the 
dynamical mass $\rm 10^8~M_{\odot}$ enclosed within a 70 pc radius region 
around the central engine (as derived both from cold and hot molecular gas 
kinematics independently) corresponds to a super massive confined object 
or a mass-distributed component (stellar cluster and molecular/dust gas). As 
there is most probably a contribution from each of these components, their
respective share is an issue we intend to address in the near future 
(Galliano \& Alloin 2001).



\acknowledgements 
We are gratefully indebted to the ESO Service Observing team on Paranal
and to the User Support Group and Archive Support Group at ESO/Garching
for efficient help. We acknowledge precious advice from C. Lidman for
the ISAAC spectra reduction and interesting comments from an anonymous 
referee.

\pagebreak


\begin{thebibliography}{}




\bibitem {} Barvainis R., Maloney P., Antonucci R., Alloin D., 1997, ApJ 484, 
695





\bibitem {} Braatz J.A.,  Wilson A.S., Gezari  D.Y. et al,
1993, ApJ 409, L5


\bibitem {} Capetti A., Axon D., Maccheto F.D. et al, 
1995a, ApJ 446, 155

\bibitem {} Capetti A., Maccheto F.D., Axon D. et al, 
1995b, ApJ 452, L87





\bibitem {} Cuby JG., Lidman C., Moutou C., Petr M., 2000, SPIE 4008, 1036
 

\bibitem {} Galliano E. \& Alloin D., 2001, in preparation









\bibitem {} Gallimore J.F., Baum, S.A., O'Dea C.P., 1997, Nature 388, 852




\bibitem {} Greenhill L.J., Gwinn C.R., 1997, Ap\&SS, 248, 261

\bibitem {} Hawarden T., Israel F., Geballe T., Wade R, 1995, MNRAS 276, 1197



\bibitem {} Kishimoto M., 1999, ApJ 518, 676




\bibitem {} Lumsden S., Moore T., Smith C. et al, 1999, MNRAS 303, 209





\bibitem {} Marco O., Alloin D., Beuzit J.L., 1997, A\&A 320, 399

\bibitem {} Marco O., Alloin D., 2000, A\&A 353, 465

\bibitem {} Moorwood A., Cuby, J.-G., Ballester P. et al, 1999, 
The Messenger, 95, 1-5


\bibitem {} Muxlow T.W., Pedlar A., Holloway A. et al, 1996, MNRAS 278, 854







\bibitem[1991] {Rota91} Rotaciuc V., Krabbe A., Cameron M. et al, 1991, ApJ, 
370, L23-L26
 
\bibitem {} Rouan D., Rigaut F., Alloin D. et al, 1998, A\&A 339, 687






\bibitem {} Schinnerer E., Eckart A., Tacconi L.J., Genzel R., Downes D., 2000,ApJ 533, 850






\bibitem {} Thatte N.,  Quirrenbach  A., Genzel R. et al, 1997,ApJ 490, 238

\bibitem[1978]{thom78} Thompson R.I., Lebofsky M.J., Rieke G.H., 1978, ApJ 222, 
L49-L53









\end{thebibliography}
\end{document}